# The Initial Mass Function: Now and Then

Harvey B. Richer† and Gregory G. Fahlman†

†*Department of Physics and Astronomy*
*University of British Columbia*
*Vancouver, British Columbia*
*Canada V6T 1Z4*
*Email: surname@astro.ubc.ca*

**Abstract.** We examine whether existing data in clusters, both old and young, and in the field of the Galactic disk and halo is consistent with a universal slope for the initial mass function (IMF). The most reasonable statement that can be made at the current time is that there is no strong evidence to support a claim of any real variations in this slope. If the IMF slope is universal then this in itself is remarkable implying that variations in metallicity, gas density or other environmental factors in the star formation process play no part in determining the slope of the mass function.

*"Current evidence favours a universal initial mass function, independent of environmental factors that can be expressed by a single power-law of slope* $x = 1$ *(Salpeter = 1.35) from* $100M_\odot$ *down to* $1M_\odot$ *and possibly even to* $0.1M_\odot$*. [1]"*

## INTRODUCTION

The mass function (MF) of a stellar population is typically expressed as $\Psi$, the number of stars / unit mass / unit volume element. The masses of the stars are not observed directly, rather it is their luminosities that are the measurable quantity so that the MF, written in terms of observables, is

$$\Psi(m) = \Theta(M_V) \cdot \frac{dM_V}{dL} \cdot \frac{dL}{dm} \qquad (1)$$

where $m$ is the mass, $\Theta(M_V)$ the luminosity function (for illustrative purposes in $M_V$), $\frac{dM_V}{dL}$ the bolometric correction and $\frac{dL}{dm}$ the mass-luminosity relation. The bolometric correction is usually obtained from theory but can be measured empirically, while the mass-luminosity relation is readily derived for population I from binary systems but currently must be obtained from theory for population II due to a lack of empirical data. Historically, this has been an important difficulty in deriving accurate MFs from luminosity functions for population II samples, but enormous progress has recently been made (see the section below on globular clusters). When the bolometric correction is known, equation 1 reduces to its simplest form

$$\Psi(m) = \Theta(M_V) \cdot \frac{dM_V}{dm} \tag{2}$$

where it is now explicitly seen that the observed luminosity function must be multiplied only by the *slope* of the $M_V$-mass relation to obtain the MF. This MF is often assumed to be a power-law so that

$$\Psi(m) \propto m^{-(1+x)}, \tag{3}$$

$x$ taking on a value of 1.35 in the work of Salpeter [2]. Salpeter derived this slope using Galactic field stars in the mass range $0.5 - 10 M_\odot$.

Moffat, quoted above, was mainly interested in young starbursts when he made his claim of a universal MF slope. We expand his discussion to include much older open clusters, the MF slope in the Galactic disk and halo, and the MF in globular clusters.

## YOUNG CLUSTERS, ASSOCIATIONS AND YOUNG FIELD STARS

Massey *et al.* [3–5] have carried out systematic ground-based work on the MF in young stellar aggregates and in the field both for the Galaxy and in the Magellanic Clouds. Similar research has been done in nearby resolved systems by Hunter *et al.* [6–9] using *HST*. These studies, together with others, point to a MF slope for all the systems, independent of which galaxy they reside in or the physical conditions in the gas out of which they formed, in the range of $x = 1 - 1.3$ with errors of 10 - 20% in the slope. In these systems it is likely that we are actually seeing the slope of the initial mass function (IMF), at least at the highest mass end, as these objects are too young to have suffered much dynamical evolution. Williams et al. [10] have extended work of this sort to very low masses using infrared observations in the star forming $\rho$ Oph cluster. They find a value of $x = 1.1$ which extends from $\sim 4 M_\odot$ down below $0.1 M_\odot$, although there is some uncertainty in establishing stellar masses from the low resolution infrared spectroscopy used in their analysis.

The one serious divergent data point in the universality of MF slopes for young stars is found in the field of the Galaxy and the Magellanic Clouds where Massey *et al.* [4] find very steep slopes, up to 4. It is likely that in this case we are *not* seeing the IMF as these stars probably recently escaped from their clusters of origin which are likely to be more efficient at retaining the most massive objects.

## OLD OPEN CLUSTERS AND LOW MASS DISK STARS

The most comprehensive work on the MFs in older Galactic clusters is due to Francic [11]. In a sample of eight clusters ranging in age from $\sim 10^8$ to $5 \times 10^9$ years, Francic found that the MFs for the 5 youngest systems were all in the range of $x = 1$ with very little scatter. In the three oldest clusters (NGC 6633, NGC 752, M67) the currently observed MF is weighted toward higher mass stars. There is little doubt that this has resulted from the loss of low mass stars through dynamical processes.

Recent efforts at deriving the MF for the disk of the Milky Way have concentrated on improving the theoretical models for low mass stars and on the data sets used to construct the luminosity functions from which the MF is constructed. New models have been presented by Méra *et al.* [12] who then used the luminosity function derived by Kroupa [13] to obtain a MF for the disk of the Galaxy. This MF is well represented by a power-law with a slope of $x \simeq 0.7$.

## THE MASS FUNCTION IN GLOBULAR CLUSTERS

There have been two recent developments which have dramatically altered our view of the MF in Galactic globular clusters. The first of these is observational. New data obtained with the Hubble Space Telescope (*HST*) are both more accurate (as they are less affected by crowding) and penetrate to fainter magnitudes than ground-based data so that the cluster luminosity functions obtained with *HST* are better defined and descend to significantly lower masses. Further, *HST*-derived luminosity functions can be secured in the crowded cores of many Galactic globular clusters so that the effects of mass segregation can be more readily quantified. These points are illustrated in Figure 1 where the color-magnitude diagram (CMD) of M4 obtained from the Las Campanas Observatory [14] is compared with an *HST* cycle 4 CMD [15].

The second major improvement in deriving MFs for globular clusters comes from some very impressive theoretical work. Both the Lyon group in France, led by Gilles Chabrier, and the group in Victoria under Don VandenBerg have been constructing new interior models for metal-poor stars with revised equations of state, and are fitting these to model atmospheres calculated by France Allard and her collaborators. These models, while not yet in final form, are a vast improvement over earlier ones. We illustrate this in Figure 2 where the published models of D'Antona and Mazzitelli [16] are compared with the CMD of M4 in the left panel of the Figure while new VandenBerg-Allard [17] models are shown on the right. It is clear that the older models do a rather poor job of fitting the lower portions of the M4 CMD whereas the more recent models are now an excellent fit all the way to the limit of the data. Significant errors will hence arise into the calculation of MFs constructed from models which are poor fits to the cluster CMDs.

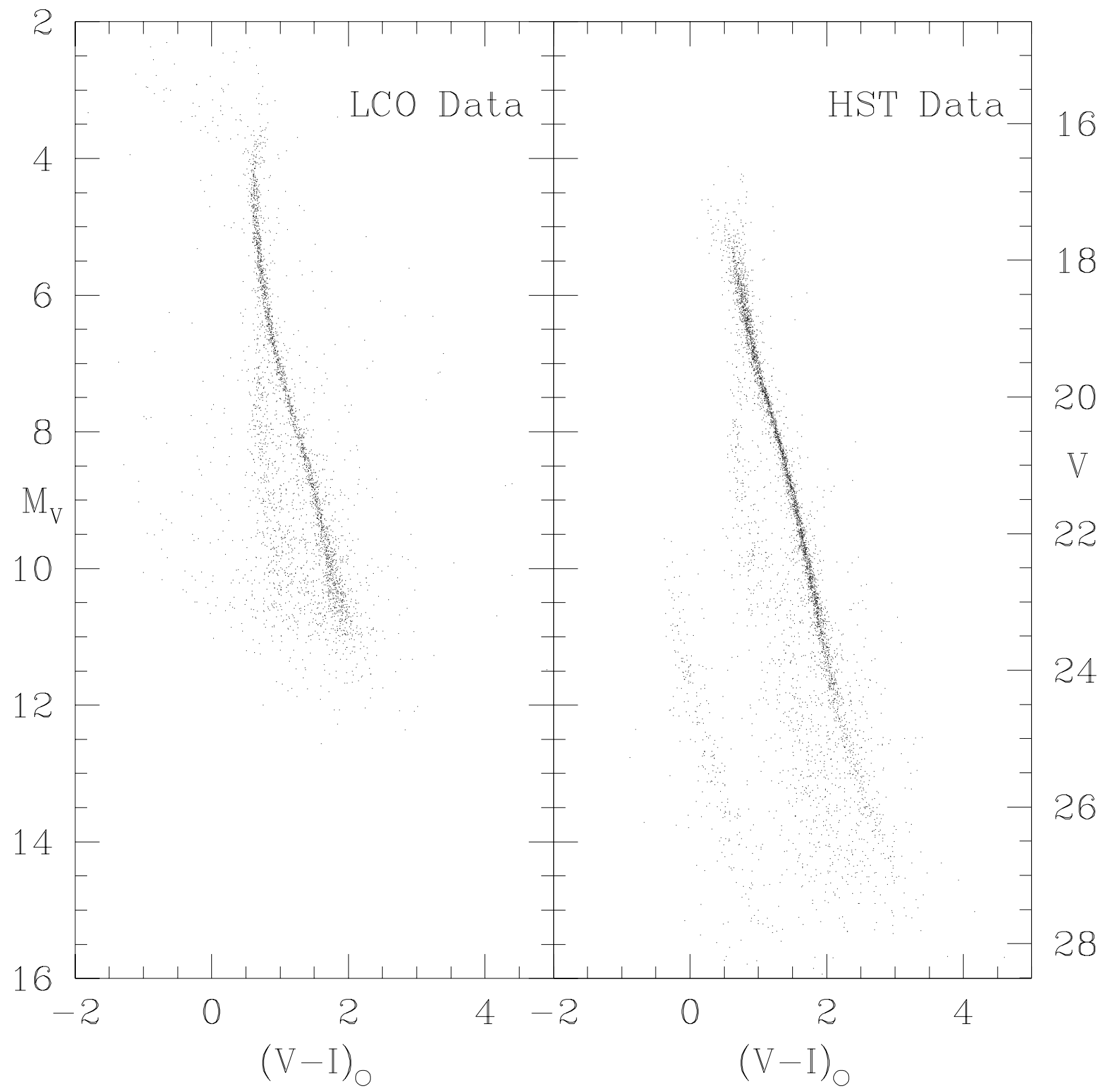

**FIGURE 1.** A ground-based CMD for the Galactic globular cluster M4 obtained from Las Campanas Observatory (left panel) compared with one secured with *HST*. Note that not only is the main sequence more tightly defined and descends to much fainter objects in the *HST* CMD, but that an extensive white dwarf cooling sequence is also discovered.

King *et al.* [18] used the D'Antona and Mazzitelli stellar models together with luminosity functions derived from *HST* observations to produce MFs for 4 globular clusters (47 Tuc, NGC 6397, M15, M30). To these they added the MF of $\omega$ Centauri taken from Elson *et al.* [19] and demonstrated that all of these (with the exception of NGC 6397) appeared to have about the same MF slope. These clusters (NGC 6397 excluded) are not expected to have lost many of their low mass stars as, according to the cluster evolutionary models of Gnedin and Ostriker [20], their destruction rates due to disk and bulge shocks and tidal effects are small. All the data were obtained near the cluster half-mass radii where the local MF well approximates the global one. Below a mass of $0.4 M_{\odot}$, where the mass segregation effects are small [21], the MF slopes for these globular clusters are in the neighborhood of $x = 0.7$ with a range of about $\pm 0.2$. These slopes may be a good representation of the cluster IMF values. The slope for NGC 6397 is near $x = 0.5$. However, what should be kept in mind are the possible errors introduced in using the D'Antona-Mazzitelli models. These produce MF slopes that are steeper than those generated using the newer VandenBerg-Allard mass-luminosity relation. For example, the MF of 47 Tuc, determined using these latter models, has a slope near $x = 0.4$ which is to be compared with $x = 0.8$ using the D'Antona-Mazzitelli models. Until the theoretical models for low mass population II stars are finalized, this level of uncertainty will remain in the transformation from luminosity functions to MFs and not until then will a clean comparison with MF slopes for population I be possible.

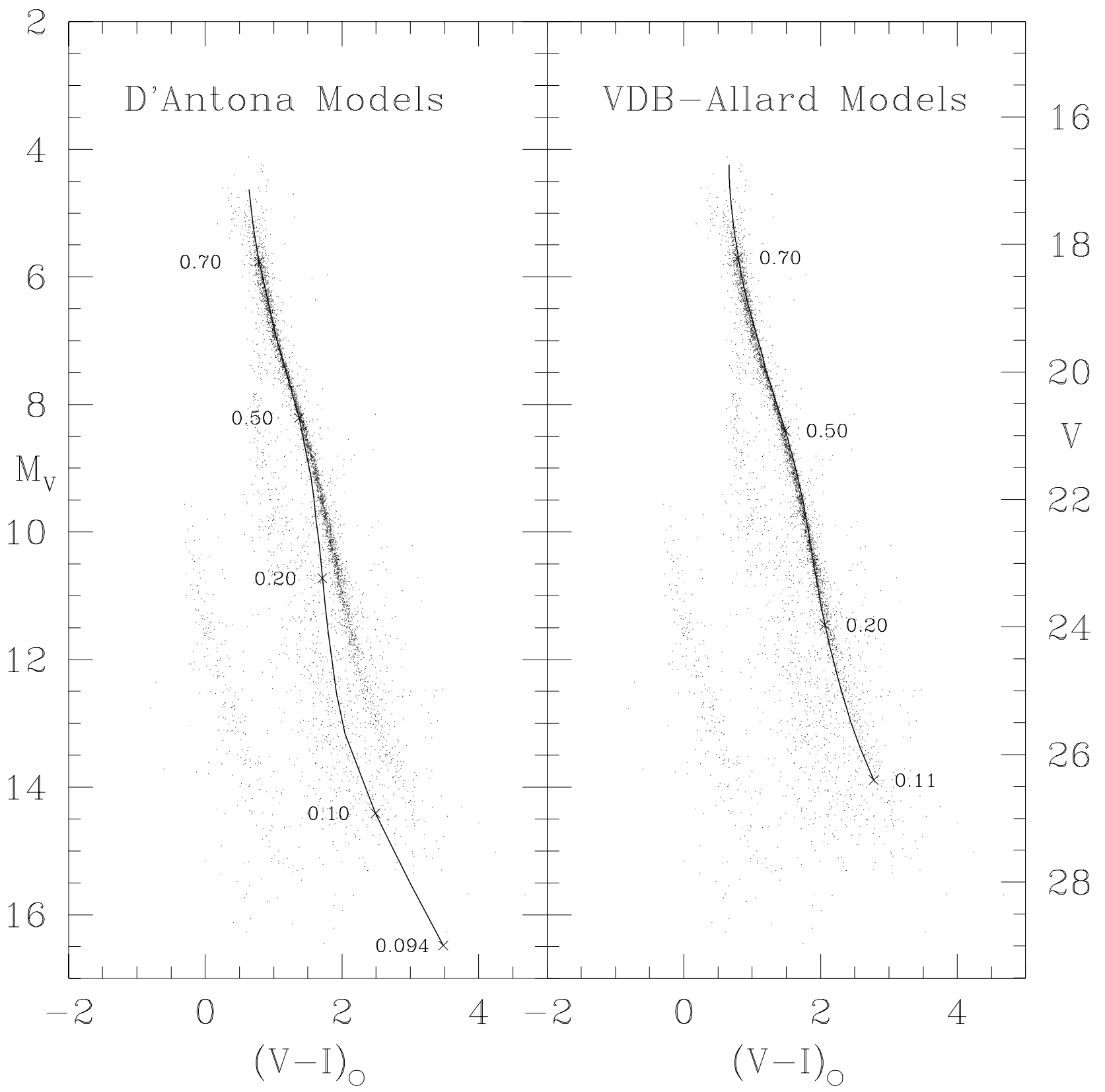

**FIGURE 2.** D'Antona and Mazzitelli 10 Gyr isochrone for $[m/H] = -1.3$ compared with the CMD of M4 in the left hand panel. The small numbers indicate the stellar masses. In the right panel new interior models calculated by VandenBerg for $[m/H] = -1.0$ are combined with model atmospheres by Allard to produce isochrones which provide a much improved fit to the data.

Recent discoveries with *HST* allow for an extension of globular cluster MFs to stars more massive than those at the turnoff. We discuss this with reference to results on M4 [22,15]. The CMD which we obtained for this cluster is displayed in Figure 1 wherein a well-populated white dwarf cooling sequence can be seen. These objects evolved from stars more massive than the current cluster turnoff so they provide the potential of extending the cluster MF to higher masses. Current turnoff stars in a cluster such as M4 ($[m/H] = -1.3$) have masses of $\sim 0.8 M_\odot$ [23]. Because counting faint white dwarfs in the environs of a crowded field of a globular cluster is subject to serious incompleteness, the oldest white dwarfs in M4 for which we are confident of the statistics are 3.7 Gyr. In this cluster, the progenitor of these stars had masses of $0.9 M_\odot$ so the present sample of white dwarfs allow us to extend the MF upward by about $0.1 M_\odot$. The observed main sequence MF of M4 is very flat, with $x \simeq -1$ [24], and the white dwarfs are found to extend this MF at about the same slope. This very flat MF is unlikely to be the cluster IMF as the orbit of M4 keeps it close to the Galactic plane [25] where shocking and tidal effects can efficiently strip low mass stars from it.

The possibility exists, however, to extend the cluster MF to much higher masses. This is discussed in detail in Richer *et al.* [15] and we outline the ideas here. The number of white dwarfs in a cluster brighter than some magnitude depends on four factors: the age of the cluster, the cooling time of the white dwarfs to that magnitude, the dependence of main sequence lifetime on stellar mass ($T_{MS} \propto m^{-\gamma}$), and the cluster IMF ($\Psi(m) \propto m^{-(1+x)}$). The last two factors work in the following way. The larger is $x$, the steeper the mass function, and the fewer the number of high mass stars that eventually become white dwarfs. On the other hand, the bigger is $\gamma$, the shorter the amount of time massive stars spend on the main sequence, and hence the quicker they produce white dwarfs. Under the assumption of no loss of stars from a cluster, the white dwarf luminosity function will then contain information on the cluster IMF for a given value of $\gamma$. Very faint white dwarf cooling sequences are required to use this technique efficiently, but such data are possible to obtain with *HST*.

## THE MASS FUNCTION OF THE GALACTIC SPHEROID AND HALO

In discussing the MF of the outer environs of the Galaxy away from the disk, a clear distinction between the Galactic spheroid and halo must be kept in mind. If stars are found in a density distribution which varies as $r^{-3.5}$ then it is the spheroid that is being examined. If the halo of the Galaxy has a stellar component associated with it, then these objects will be distributed as $r^{-2}$ so as to account for the flat rotation curve of the Galaxy. It is for this reason that not only should the luminosity function (or MF) of outer Galaxy stars be measured, but, it is critical to examine the density distribution simultaneously so as to be sure which component of the Galaxy is being explored.

With this idea in mind, Richer and Fahlman [26] (hereafter RF) initiated a program using CFHT of determining the MF of stars in the outer Galaxy from stars found *in situ.* The major difficulties with this approach are star/galaxy separation at the faint limits and the assignment of stars to the disk, to the thick disk and to the spheroid or halo. In a single $49\square'$ field with a sample of only 31 stars, RF found a very steep MF slope ($x = 3.5$) with the distribution of the stars well represented by $\rho \propto r^{-3.5}$. In a second, similarly-sized field (unpublished), a somewhat flatter MF slope ($x = 2.0$) was measured, and, again, the density distribution fit that expected for the spheroid. These large differences in the derived MF slopes may be suggesting that the outer parts of the Galaxy are quite inhomogeneous as would be expected if it formed from a number of accretion events.

The results of RF were analyzed in some detail by Reid *et al.* [27] who pointed out that the observed MF was likely to have been contaminated by stars in the Galactic thick disk. This could produce an artificial upturn at the low mass end and thus a spuriously rapidly rising slope. Reid *et al.* suggested that all stars redder than $(V - I) = 1.75$ observed by RF actually belong to the thick disk. To avoid the possible inclusion of thick disk objects, we have recomputed the observed MF eliminating all stars below a Z-distance of 5 kpc. Excluding these stars from the original RF sample leaves only 23 objects, while removing them from the second field produces a sample of 45 stars. The MF derived from these very small groups of stars is much flatter with $x \simeq 0.7$, but the error in this slope is large. The objects still fit a spheroid density distribution reasonably well with $\rho \propto r^{-2.9}$.

An alternate approach to deriving the MF for the outer regions of the Galaxy is to study a sample of nearby high proper-motion stars. This technique suffers from the possibility of bias in the selection of the stars which is somewhat difficult to correct for. Further, with such a local sample, it is not possible to solve for the density distribution of the stars. Dahn *et al.* [28] have derived a luminosity function for such objects and we use the new VandenBerg-Allard models to convert this to a MF. This function is displayed in Figure 3 where the approximately linear part of the diagram, from $\log(M/M_\odot) = -0.4$ to $-0.8$, has a slope of $x = 0.6$ and the MF clearly appears to turn over below a mass of $\sim 0.16 M_\odot$ which is well above the hydrogen-burning limit for metal-poor stars. If this MF represents that of the Galactic halo satisfactorily, with no upturn at lower masses, then very low mass stars and brown dwarfs are unimportant contributors to the total mass budget of the Galaxy. This conclusion is also in agreement with recent results from the microlensing experiments [29].

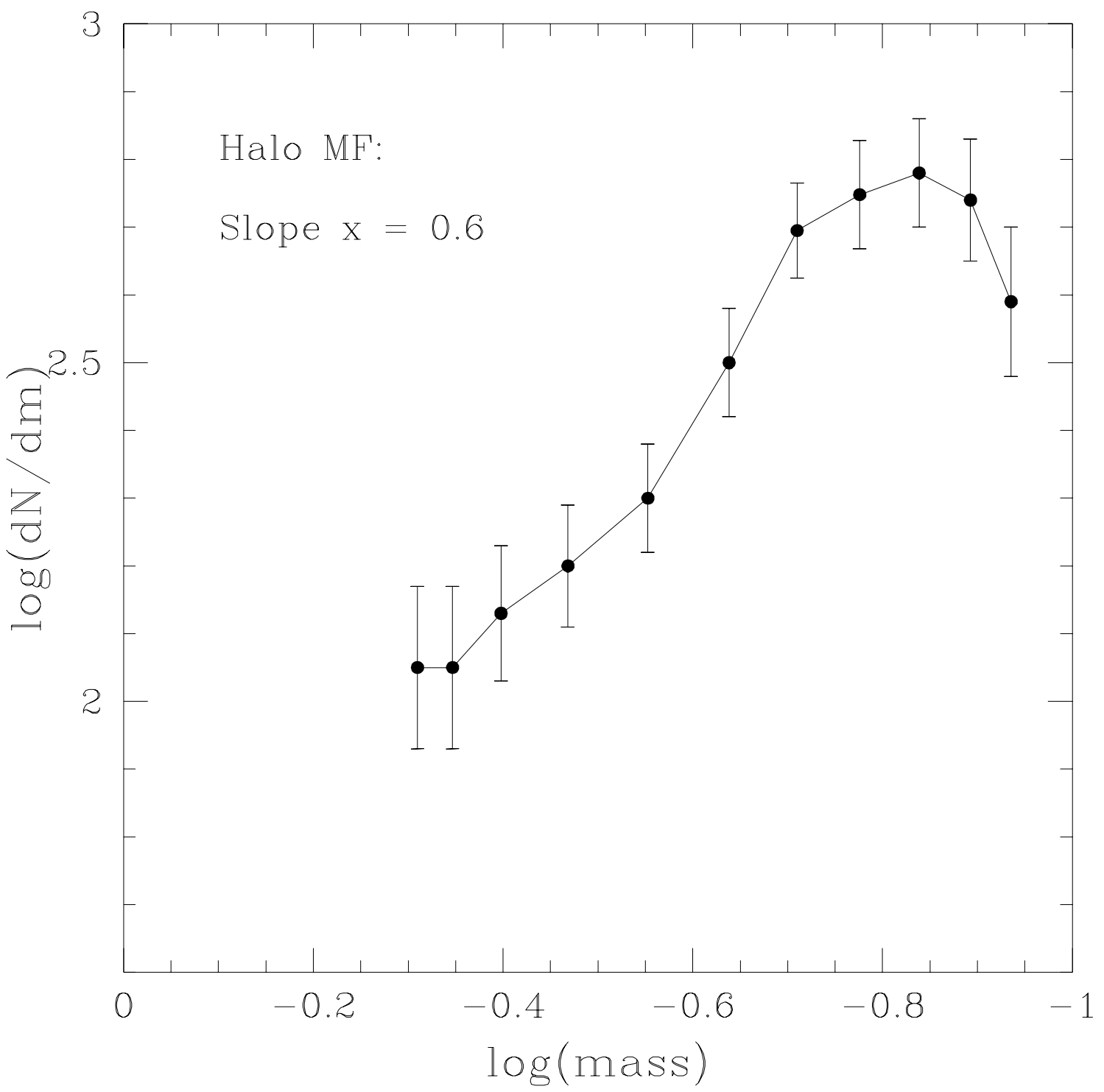

**FIGURE 3.** Mass function for high-velocity stars in the solar neighborhood. The data are from Dahn *et al.* and the models used to convert the luminosity function into a mass function are from Vandenberg and Allard.

## A SUMMARY OF THE MASS FUNCTIONS

To illustrate the similarity of the MFs in extremely different environs, we combine a representative sample of them into a single diagram below (Figure 4). Vertical offsets were applied to each data set to separate the MFs so that only the slopes and the position along the mass axis for the different samples are of interest. The similarity of the slopes in this diagram is remarkable considering the widely different environments under which these stars have formed. There may be some tendency for the MFs involving older stars to have $x$ somewhat smaller, but given the current uncertainty in the modelling, it is premature to attach much weight to this. Also of interest is the mass coverage of each MF, in particular, the manner in which the old open clusters in the Galaxy fill the space between the OB associations and the disk. The average MF slope for all the populations shown is $x = 0.86 \pm 0.23$, somewhat flatter than the Salpeter [2] slope.

Understanding, theoretically, the universality of MF slopes will surely be a challenge. A clue might be found, however, in the precursors to the stars themselves. In regions where star formation is not yet underway, a number of observations have shown that the spectrum of clump masses are already well fitted by a power law with slope near $x = 1$ [30]. Turbulence in the gas may then be invoked as the cause of this spectrum, but the origin of the turbulence in regions where hot stars do not as yet exist remains a puzzle.

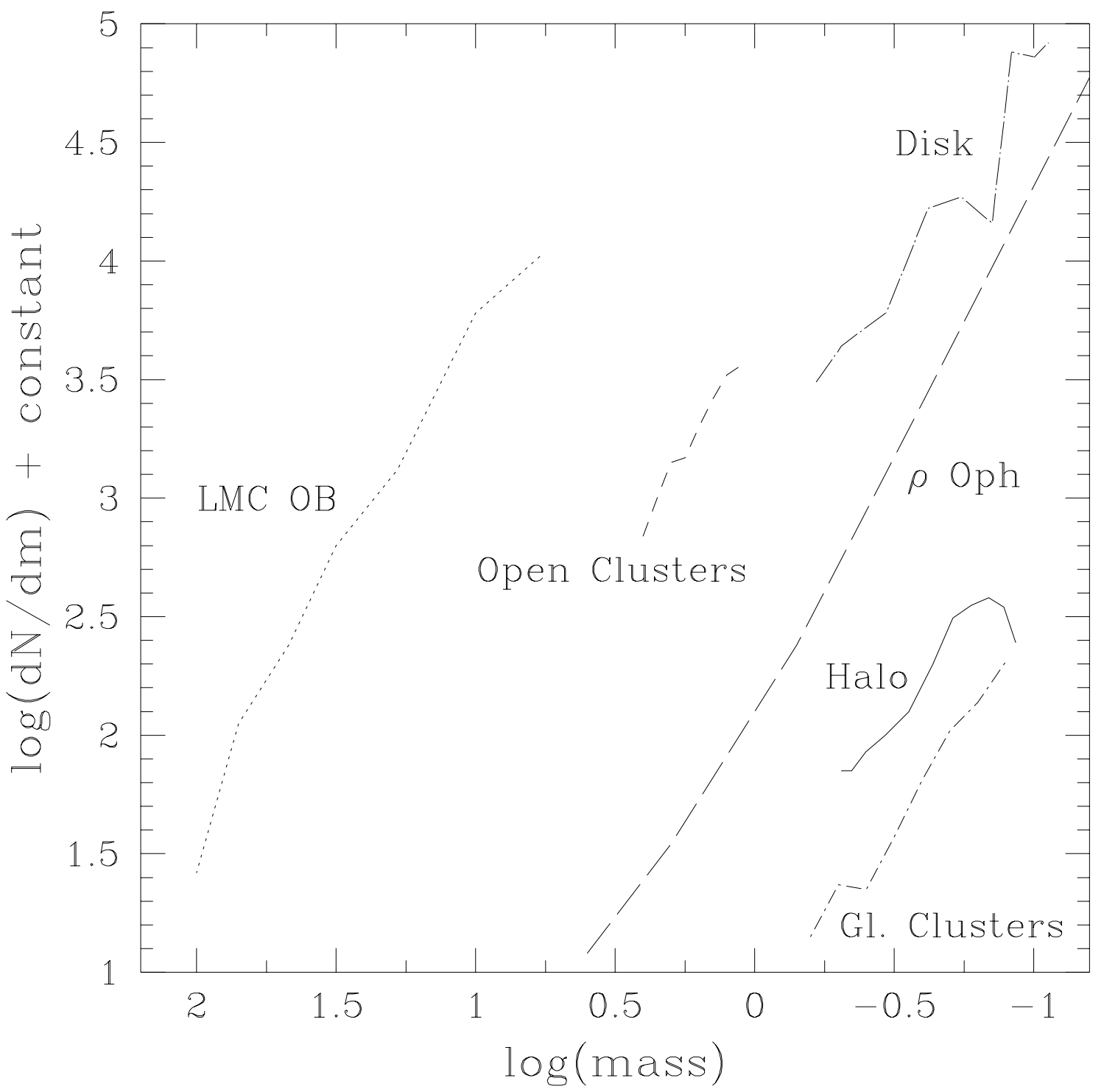

**FIGURE 4.** Mass functions in a wide variety of environments from extremely young OB associations in the LMC, to a selection of Galactic open clusters and the Galactic disk, through to metal poor globular clusters (M15, M30) and the Galactic halo. Arbitrary offsets have been applied to the data so that only the slopes and the coverage in mass are of interest in this Figure. Sources of the data can be found in the relevant sections in the text.